\documentclass[10pt,conference,letterpaper]{IEEEtran}
\usepackage{graphicx}
\usepackage{enumitem}
\usepackage{latexsym}
\usepackage{amsmath}
\usepackage{amssymb}
\usepackage[normalem]{ulem}
\usepackage{comment}
\usepackage{cite}
\usepackage{bm}
\usepackage{url}
\usepackage{xcolor}

\newtheorem{proposition}{\hspace{-3mm}Proposition}
\newtheorem{corollary}{\hspace{-3mm}Corollary}
\newtheorem{theorem}{\hspace{-3mm}Theorem}
\newtheorem{definition}{\hspace{-3mm}Definition}
\newtheorem{remark}{\hspace{-3mm}Remark}

\begin{document}

\title{Three-Pass Identification Scheme Based on MinRank Problem with Half Cheating Probability}

\author{\IEEEauthorblockN{Bagus Santoso}
	\IEEEauthorblockA{%
		Department of Computer and Network Engineering, \\
		The University of Electro-Communications, \\
		Email: \url{santoso.bagus@uec.ac.jp}}
	\and
	\IEEEauthorblockN{Yasuhiko Ikematsu}
	\IEEEauthorblockA{%
		Institute of Mathematics for Industry,
		Kyushu University \\
	}
	%
	\and
	\IEEEauthorblockN{Shuhei Nakamura}
	\IEEEauthorblockA{%
		Department of Liberal Arts and Basic Sciences,
		Nihon University \\
	}
	\and
	\IEEEauthorblockN{Takanori Yasuda}
	\IEEEauthorblockA{%
		Institute for the Advancement of Higher Education, \\
		Okayama University of Science \\
	}
}


\maketitle

\begin{abstract}
	In Asiacrypt 2001, Courtois proposed the first three-pass zero-knowledge
	identification (ID) scheme based on the MinRank problem. However,
	in a single round of Courtois' ID scheme, the cheating probability, i.e., the success
	probability of cheating prover, is 2/3, which is larger than half.
	Although Courtois also proposed a
	variant scheme which is claimed to have half cheating probability,
	its security is not formally proven and it
	requires another hardness assumption on a specific one-way function and
	that verifier always generates challenges according to a specific
	non-uniform distribution.

	In this paper, we propose the first three-pass zero-knowledge ID scheme
	based on the MinRank problem with the cheating probability of exactly
	half for each round, even with only two-bit challenge space,
	\underline{without any additional assumption}.
	Our proposed ID scheme requires less
	number of rounds and less total
	communications costs compared to Curtois' under the same
	security level against impersonation.
\end{abstract}





\section{Introduction}
In 1997, P.\! Shor~\cite{DBLP:journals/siamcomp/Shor97} showed polynomial-time quantum algorithms to break integer factoring 
and discrete logarithm based cryptosystems.
Therefore, 
we need to develop cryptosystems having a resistance to quantum computer attacks.
The research area to study such cryptosystems is 
called post quantum cryptography (PQC)~\cite{PQC}. 
The most promising candidates for PQC are 
based on lattice, isogeny, coding theory, and multivariate 
polynomial problems.
In particular, one of computational problems based on multivariate polynomials is 
\emph{multivariate quadratic}
(MQ)
problem, which finds a solution to a system of quadratic equations over a finite field.
In general, MQ problem is the foundation for constructing
multivariate public key cryptosystems (MPKC).
 \;There have been a lot of multivariate schemes, 
HFE \cite{DBLP:conf/eurocrypt/Patarin96}, UOV \cite{DBLP:conf/eurocrypt/KipnisPG99}, Rainbow \cite{DBLP:conf/acns/DingS05}, and so on.
Among them, Rainbow was chosen as a third round candidate \cite{NIST_Rainbow} in NIST PQC standardization project~\cite{NIST}.
  
It is known that 
many cryptanalysis against 
multivariate schemes (including Rainbow)  are not only based on 
MQ problem, but also another computational problem called MinRank problem.
MinRank problem 
is the problem of finding a linear combination
with a specified rank from a given finite set of matrices.
MinRank problem is proven to be 
NP-complete \cite{DBLP:journals/jcss/BussFS99}.
Therefore, we can consider cryptographic schemes based on this problem.

In 2001, 
Courtois~\cite{DBLP:conf/asiacrypt/Courtois01} 
proposed the first three-pass zero-knowledge
identification (ID) scheme
based on MinRank problem. 
In this ID scheme, the cheating probability,   
i.e., the success probability of cheating prover, is 2/3, 
which is larger than half (=1/2).
As a result, to achieve the desired security level
against impersonation, Courtois'  ID scheme needs to be 
repeated in larger number of rounds compared to 
the common ID scheme with half cheating probability
such as Feige-Fiat-Shamir~\cite{jofc-1988-14085} or 
Schnorr~\cite{DBLP:journals/joc/Schnorr91} 
ID schemes.
This makes the total communication cost of Courtois' ID scheme
quite high in practice.
In the same paper~\cite{DBLP:conf/asiacrypt/Courtois01}, Courtois also proposed a variant  
of his ID scheme,
and claimed 
that the variant  has half cheating probability 
by employing  \underline{additional assumptions} as follows:  
(1) the verifier sends the challenge according to a certain
fixed distribution and
(2) a certain special function satisfies one-wayness.
However, Courtois did not provide any formal proof 
that the variant scheme is secure. 
Moreover, it is not clear how the variant scheme  
will maintain privacy against an adversary which acts 
as a malicious verifier where it sends challenge 
according to arbitrary distribution.

In this paper, we propose a new three-pass ID scheme based on MinRank problem.
By assuming the hardness of decisional MinRank problem and the existence of 
perfectly hiding and computational binding commitment,
without using any additional assumption, we can prove that 
the probability that an adversary without a valid secret key 
being accepted by the adversary is at most half ($=$1/2).
Hence, the number of rounds which are needed  for our proposed scheme 
to achieve the desired security level is less than Courtois' ID scheme.

This paper is organized as follows. 
In Section 2 we explain the MinRank problem.
In Section 3 we describe our ID scheme and its security properties.
In Section 4 we prove the theorems related to 
the properties of our scheme.
In Section 5 we discuss the selection of practical parameters.
Finally, we conclude our paper in Section 6.

\section{Preliminaries}

\paragraph*{Notations and Consensus. }
Unless noted otherwise, let any algorithm in this paper be a probabilistic
polynomial time Turing Machine. 
We also define as follows: let $\mathbb{F}$ be a finite field used throughout this paper, 
$\mathrm{M}_n(\mathbb{F})$ be the set of $n$-square matrices over $\mathbb{F}$ and
$\mathrm{GL}_n(\mathbb{F})$ be the set of $n$-square invertible matrices over $\mathbb{F}$.

\begin{definition}
	[Search MinRank Problem]\label{def:Search-Minrank-Problem} The
	search MinRank problem is defined as follows. Given a positive integer
	$r\in\mathbb{N}$ and $m$ matrices
	$M_{0},M_{1,}\ldots,M_{m-1}\in\mathrm{M}_n(\mathbb{F})$, 
    find $\alpha=(\alpha_{1},\ldots,\alpha_{m-1})\in\mathbb{F}^{m-1}$
	such that $\mathrm{rank}(M)=r$, where $M=\sum_{i=1}^{m-1}\alpha_{i}M_{i}-M_{0}$.
\end{definition}

\subsection*{Decisional MinRank Problem}

In this paper, we use the hardness of the decisional version of the
MinRank problem as the basic assumption of the security since it is
much simpler to prove the security based on the decisional version
compared to the search version above.

\begin{definition}
	[Decisional MinRank Problem] An algorithm $\mathcal{D}$ is said
	to $(t,\varepsilon)$-solve the decisional MinRank problem associated
	with the finite field $\mathbb{F}$ and $r,m,n\in\mathbb{N}$ if $\mathcal{D}$
	runs in $t$ units of time and the following holds.

	\begin{align*}
		 & \left| %
		\Pr\left[\mathcal{D}^{\mathsf{IGen}}(\mathbb{F},r,m,n)=1\right]
		-
		\Pr\left[\mathcal{D}^{\mathsf{LossyGen}}(\mathbb{F},r,m,n)=1\right]
		\right|
		\geqq\varepsilon,
	\end{align*}
	where:
	\begin{itemize}
		\item $\mathcal{D}^{\mathsf{IGen}}$ denotes that $\mathcal{D}$ receives
		      the input from the oracle $\mathsf{IGen}$ which generates an instance
		      of MinRank problem that has at least one solution, i.e., $m$  
              matrices: $M_{0},M_{1,}\ldots,M_{m-1}\in\mathrm{M}_n(\mathbb{F})$,
		      such that there exists $\alpha=(\alpha_{1},\ldots,\alpha_{m-1})\in\mathbb{F}^{m-1}$
		      satisfying the following:

		      \begin{align}
			      \mathrm{rank}\left(\sum_{i=1}^{m-1}\alpha_{i}M_{i}-M_{0}\right) & =r,\label{eq:Igen}
		      \end{align}

		\item $\mathcal{D}^{\mathsf{LossyGen}}$ denotes that $\mathcal{D}$ receives
		      the input from the oracle $\mathsf{LossyGen}$ who generates $m$
              random matrices $M_{0},M_{1,}\ldots,M_{m-1}\in\mathrm{M}_n(\mathbb{F})$,
              which \uline{do not} necessarily
		      have $\alpha=(\alpha_{1},\ldots,\alpha_{m-1})\in\mathbb{F}^{m-1}$
		      satisfying Eq. (\ref{eq:Igen}).
	\end{itemize}
	The decisional MinRank problem associated with the finite field $\mathbb{F}$
	and $r,m,n\in\mathbb{N}$ is said to be $(t,\varepsilon)$-hard if
	there is no algorithm $\mathcal{D}$ which $(t,\varepsilon)$-solves
	the problem.
\end{definition}
\begin{remark}
	Buss et al. \cite{DBLP:journals/jcss/BussFS99} and  
    Courtois\cite{DBLP:conf/asiacrypt/Courtois01}
	have proven that the decisional MinRank problem is NP-Complete.
\end{remark}

\section{Proposed Scheme }

In this section, first we describe our proposed identification scheme.
Then we show that our proposed scheme satisfies the standard properties
such as completeness, soundness, and zero-knowledgeness.

\subsection{Construction}

\subsubsection*{Key Generation}

Given the security parameter as input, the key generator generates
the public $pk$ and the secret key $sk$ which satisfy  the following properties.
The public key $pk$ consists of a positive integer $r\in\mathbb{N}$
and $m$ matrices $M_{0},M_{1,}\ldots,M_{m-1}\in\mathrm{M}_n(\mathbb{F})$.
The secret key $sk$ consists of $\alpha=(\alpha_{1},\ldots,\alpha_{m-1})\in\mathbb{F}^{m-1}$
such that $\mathrm{rank}(M)=r$, where $M=\sum_{i=1}^{m-1}\alpha_{i}M_{i}-M_{0}$.

\subsubsection*{Interactive Protocol}

A single elementary round of interactive protocol between a prover
$P(pk,sk)$ and a verivier $V(pk)$ is described as follows.
Similar to Courtois' ID scheme \cite{DBLP:conf/asiacrypt/Courtois01},
we also employ the hash function $H$ which acts as
a commitment with \emph{perfectly hiding} and
\emph{computational binding} properties.

\begin{itemize}[leftmargin=15mm]
	\item [{Step{ }1:}] $P$ randomly generates $S_{0},S_{1},T_{0},T_{1}\in\mathrm{GL}_n(\mathbb{F})$
	      and $X_{0},X_{1}\in\mathrm{M}_n(\mathbb{F})$.
	      Next, $P$ randomly generates $\beta_{0}=(\beta_{0,1},\ldots,\beta_{0,m-1})\in\mathbb{F}^{m-1}$ and
	      $\beta_{1}=(\beta_{1,1},\ldots,\beta_{1,m-1})\in\mathbb{F}^{m-1}$, then 
	      computes the following:

	      \hspace{-10mm}
	      \begin{tabular}{l@{\hspace{-1mm}}|@{\hspace{-2mm}}l}%
		      \begin{minipage}[t]{43mm}\vspace{-5mm}
			      \begin{gather*}
				      N_{0} =\sum_{i=1}^{m-1}\beta_{0,i}M_{i} \\
				      U_{0,0}  =T_{0}N_{0}S_{0}+X_{0} \\
				      U_{0,1} = T_{0}MS_{0}+U_{0,0} \\
				      R_{0} = (S_{0},T_{0},X_{0})
			      \end{gather*}
		      \end{minipage}
		       &
		      \begin{minipage}[t]{43mm}\vspace{-5mm}
			      \begin{gather*}
				      N_{1}=\sum_{i=1}^{m-1}\beta_{1,i}M_{i}\\
				      U_{1,0}=T_{1}N_1S_{1}+X_1 \\
				      U_{1,1}=T_{1}MS_{1}+U_{1,0} \\
				      R_{1}=(S_{1},T_{1},X_{1})
			      \end{gather*}
		      \end{minipage}
	      \end{tabular} \\

	      Finally, $P$ sends $Y=(Y_{0},Y_{1})$ to $V$ where the followings hold.
	      %
	      \begin{align}
		      Y_{0} & =(H(U_{0,0}),H(U_{0,1}),H(R_{0}))  \\
		      Y_{1} & =(H(U_{1,0}),H(U_{1,1}),H(R_{1})).
	      \end{align}

	\item [{Step{ }2:}] $V$ parses $Y_{0}$ and $Y_{1}$ as $Y_{0}=(Y_{0,0},Y_{0,1},Y_{0,2})$
	      and $Y_{1}=(Y_{1,0},Y_{1,1},Y_{1,2})$. Then, $V$ chooses randomly
	      $c\in\left\{ 0,1,2,3\right\} $ and sends $c$ to $P$.
	\item [{Step{ }3:}] $P$ computes $Z_{0,0},Z_{0,1},Z_{1,0},Z_{1,1}$ according
	      to the value of $c$ as follows. \\
	      \ \\
	      \begin{tabular}{l|cc}
		      \hline
		      \textbf{Case} $c=0$:
		       & $Z_{0,0}=U_{0,0}$,         & $Z_{1,0}=R_{1}$,            \\
		       & $Z_{0,1}=U_{0,1}$,         & $Z_{1,1}=\beta_{1}$.        \\
		      \hline
		      \textbf{Case} $c=1$:
		       & $Z_{0,0}=R_{0}$,           & $Z_{1,0}=R_{1}$,            \\
		       & $Z_{0,1}=\beta_{0}$,       & $Z_{1,1}=\beta_{1}+\alpha$. \\
		      \hline
		      \textbf{Case} $c=2$:
		       & $Z_{0,0}=R_{0}$,           & $Z_{1,0}=R_{1}$,            \\
		       & $Z_{0,1}=\beta_{0}+\alpha$ & $Z_{1,1}=\beta_{1}$         \\
		      \hline
		      \textbf{Case} $c=3$:
		       & $Z_{0,0}=R_{0}$,           & $Z_{1,0}=U_{1,0}$,          \\
		       & $Z_{0,1}=\beta_{0}$,       & $Z_{1,1}=U_{1,1}$.          \\
		      \hline
	      \end{tabular}
	      \\
	\item [{Step{ }4:}] %
	      $V$ parses $Z=(Z_{0},Z_{1}$) into $Z_{0,0},Z_{0,1},Z_{1,0},Z_{1,1}$.
	      And then $V$ performs verification procedure according to the value
	      of $c$ as shown in Fig. \ref{fig:checking}.
	      If all corresponding checking equations hold, $V$ outputs $1$ (accept),
	      otherwise $V$ outputs $0$ (reject).
	      %
	      %
\end{itemize}

\begin{figure*}[htbp]
	\fbox{
		\scalebox{0.75}{%
			\begin{tabular}{l|l}
				\begin{minipage}[t]{1.3\columnwidth}%
					\begin{description}
						\item [{Case{ }$c=0$:}]\  \\
						      $Z_{1,0}$ is parsed as $Z_{1,0}=(\widetilde{S},\widetilde{T},\widetilde{X})$
						      and $Z_{1,1}$ is parsed as $Z_{1,1}=(\widetilde{\gamma}{}_{1},\ldots,\widetilde{\gamma}_{m-1})$.
						      \begin{itemize}
							      \item []$H(Z_{0,0})\stackrel{?}{=}Y_{0,0},\quad H(Z_{0,1})\stackrel{?}{=}Y_{0,1}$, 
							            $\mathrm{rank}(Z_{0,1}-Z_{0,0})\stackrel{?}{=}r$,
							      \item []$\widetilde{S}\stackrel{?}{\in}\mathrm{GL}_n(\mathbb{F})$, $\widetilde{T}\stackrel{?}{\in}\mathrm{GL}_n(\mathbb{F})$,
							            $H(Z_{1,0})\stackrel{?}{=}Y_{1,2},\quad H(\widetilde{U})\stackrel{?}{=}Y_{1,0}$,
							            where\vspace{-2mm}
						      \end{itemize} \vspace{1mm}
						      \hspace{10mm}
						      $
							      \begin{aligned}
								      \widetilde{U}
								      =\widetilde{T}\left(\sum_{i=1}^{m-1}\widetilde{\gamma}_{i}M_{i}\right)\widetilde{S}+\widetilde{X}.
							      \end{aligned}
						      $
					\end{description}
					\begin{description}
						\item [{Case{ }$c=1$:}]\ \\
						      $Z_{0,0}$ is parsed as $Z_{0,0}=(\widehat{S},\widehat{T},\widehat{X})$
						      and $Z_{0,1}$ is parsed as $Z_{0,1}=(\widehat{\gamma}_1,\ldots\widehat{\gamma}_{m-1})$.
						      $Z_{1,0}$ is parsed as $Z_{1,0}=(\widetilde{S},\widetilde{T},\widetilde{X})$
						      and $Z_{1,1}$ is parsed as $Z_{1,1}=(\widetilde{\mu}_{1},\ldots\widetilde{\mu}_{m-1})$.
						      \begin{itemize}
							      \item []
							            $\widehat{S}\stackrel{?}{\in}\mathrm{GL}_n(\mathbb{F})$, $\widehat{T}\stackrel{?}{\in}\mathrm{GL}_n(\mathbb{F})$,
							            $H(Z_{0,0})\stackrel{?}{=}Y_{0,2}, H(\widehat{U})\stackrel{?}{=}Y_{0,0}$,
							            where\vspace{-2mm}
						      \end{itemize}
						      \begin{gather}
							      \widehat{U} =\widehat{T}\left(\sum_{i=1}^{m-1}\widehat{\gamma}_{i}M_{i}\right)\widehat{S}+\widehat{X}.
							      \label{eq:widehat-U}
						      \end{gather}

						      \begin{itemize}
							      \item[]   $\widetilde{S}\stackrel{?}{\in}\mathrm{GL}_n(\mathbb{F})$,
								      $\widetilde{T}\stackrel{?}{\in}\mathrm{GL}_n(\mathbb{F})$,
								      $H(Z_{1,0})\stackrel{?}{=}Y_{1,2}$,
								      $H(\widetilde{W}-\widetilde{T}M_{0}\widetilde{S})\stackrel{?}{=}Y_{1,1}$,
								      where\vspace{-2mm}
						      \end{itemize}
						      \begin{gather}
							      \widetilde{W} =\widetilde{T}\left(\sum_{i=1}^{m-1}\widetilde{\mu}_{i}M_{i}\right)\widetilde{S}+\widetilde{X}.
							      \label{eq:widetilde-W}
						      \end{gather}
					\end{description}
				\end{minipage}
				 &
				\begin{minipage}[t]{1.3\columnwidth}%
					\begin{description}
						\item [{Case{ }$c=2$:}]\ \\
						      $Z_{0,0}$ is parsed as $Z_{0,0}=(\widehat{S},\widehat{T},\widehat{X})$
						      and $Z_{0,1}$ is parsed as $Z_{0,1}=(\widehat{\mu}_{1},\ldots\widehat{\mu}_{m-1})$.
						      $Z_{1,0}$ is parsed as $Z_{1,0}=(\widetilde{S},\widetilde{T},\widetilde{X})$
						      and $Z_{1,1}$ is parsed as $Z_{1,1}=(\widetilde{\gamma}_{1},\ldots\widetilde{\gamma}_{m-1})$.
						      \begin{itemize}
							      \item []
							            $\widehat{S}\stackrel{?}{\in}\mathrm{GL}_n(\mathbb{F})$, $\widehat{T}\stackrel{?}{\in}\mathrm{GL}_n(\mathbb{F})$,
							            $H(Z_{0,0})\stackrel{?}{=}Y_{0,2}$,
							            $H(\widehat{W}-\widehat{T}M_{0}\widehat{S})\stackrel{?}{=}Y_{0,1}$,
							            where
						      \end{itemize}
						      \begin{gather}
							      \widehat{W}  =\widehat{T}\left(\sum_{i=1}^{m-1}\widehat{\mu}_{i}M_{i}\right)\widehat{S}+
							      \widehat{X}.
							      \label{eq:widehat-W}
						      \end{gather}

						      \begin{itemize}
							      \item []
							            $\widetilde{S}\stackrel{?}{\in}\mathrm{GL}_n(\mathbb{F})$,
							            $\widetilde{T}\stackrel{?}{\in}\mathrm{GL}_n(\mathbb{F})$.
							            $H(Z_{1,0})\stackrel{?}{=}Y_{1,2},$
							            $H(\widetilde{U})\stackrel{?}{=}Y_{1,0}$,
							            where
						      \end{itemize}
						      \begin{gather}
							      \widetilde{U} =\widetilde{T}\left(\sum_{i=1}^{m-1}\widetilde{\gamma}_{i}M_{i}\right)\widetilde{S}+\widetilde{X}.
							      \label{eq:widetilde-U}
						      \end{gather}
					\end{description}
					%

					%

					\begin{description}
						\item [{Case{ }$c=3$:}]\ \\
						      $Z_{0,0}$ is parsed as $Z_{0,0}=(\widehat{S},\widehat{T},\widehat{X})$
						      and $Z_{0,1}$ is parsed as $Z_{0,1}=(\widehat{\gamma},\ldots\widehat{\gamma}_{m-1})$.
						      \begin{itemize}
							      \item []
							            $\widehat{S}\stackrel{?}{\in}\mathrm{GL}_n(\mathbb{F})$,
							            $\widehat{T}\stackrel{?}{\in}\mathrm{GL}_n(\mathbb{F})$,
							            $H(Z_{0,0})\stackrel{?}{=}Y_{0,2}$,
							            $H(\widehat{U})\stackrel{?}{=}Y_{0,0}$,
							            where
						      \end{itemize}
						      $\qquad\qquad
							      \begin{aligned}
								      \widehat{U} =\widehat{T}\left(\sum_{i=1}^{m-1}\widehat{\gamma}_{i}M_{i}\right)\widehat{S}+\widehat{X}.
							      \end{aligned}
						      $

						      \begin{itemize}
							      \item []$H(Z_{1,0})\stackrel{?}{=}Y_{1,0},\quad H(Z_{1,1})\stackrel{?}{=}Y_{1,1}$,
							            $\mathrm{rank}(Z_{1,1}-Z_{1,0})\stackrel{?}{=}r$,
						      \end{itemize}
					\end{description}
				\end{minipage} \\
			\end{tabular}
		}
	}
	\caption{Checking equations performed by the verifier ($V$) in Step 4 of the elementary round of interactive protocol.}
	\label{fig:checking}
\end{figure*}
\begin{remark}
	The response $Z$ is said to be a \emph{valid response with respect to challenge}
	$c$ if all checking equations in the verifier side corresponding
	to the value of $c$ hold.
\end{remark}

\
\begin{remark}
	A full identification scheme consists of $\ell$ repetitions of the
	single elementary round of interactive protocol and the verifier will
	accept the prover if and only if $V$ outputs $1$ in all $\ell$
	rounds.
\end{remark}

\begin{remark}
	Here we assume that the length of the input into the hash function $H$
	is larger than that of the output, that is why we can assume that
	$H$ acts as a perfectly hiding commitment.
	We also assume that $H$ is collision resistant, i.e., for any
	polynomial algorithm, it is hard to find two distinct inputs
	with the same output.
	That is why we can assume that $H$ acts a computational binding
	commitment.
	Any common standard hash functions such as SHA-128, SHA-256, SHA-512
	is assumed to have these properties.
\end{remark}

\subsection{Completeness}

Here we show that any prover who possesses the secret key and follows
the procedure of the honest prover will always be accepted by the
verifier.
\begin{theorem}[Completeness]
	\label{thm:completeness}Let $P$ be a prover who possesses the secret
	key $sk$ corresponding to the public key $pk$ of our proposed identification
	scheme. Let $P$ generate $Y$ in Step 1 according to the described
	procedure and send it to the verifier. Then for any received challenge
	$c\in\{0,1,2,3\}$ from the verifier, if $P$ computes $Z$ according
	to described procedure, $Z$ is a valid response with respect to challenge
	$c$.
\end{theorem}

In order to prove the above theorem, it is sufficient to show that
for each challenge $c\in\{0,1,2,3\}$, $Z$ which is generated accordingly
in the procedure of the prover will satisfy all the corresponding
checking equations on the verifier side.
See Section \ref{subsec:Proof-of-Completeness}
for the detailed proof.

\subsection{Soundness }

In order to prove the soundness of our proposed scheme, we will use
the following proposition.
\begin{proposition}%
	\label{prop:three-combinations}
	Let $Y$ denote the value sent by
	the prover in the Step 1 to the verifier and let $Z^{(c)}$ denote
	the valid response with respect to the challenge $c\in\{0,1,2,3\}$.
	Then, from $Y$ and any three combinations of elements from the set
	$\{Z^{(0)},Z^{(1)},Z^{(2)},Z^{(3)}\}$ we can efficiently compute
	the solution of the search MinRank problem represented by the public key.
\end{proposition}

We describe the detailed proof of above proposition in Section \ref{subsec:Proof-of-Three-Combinations}.
Based on above proposition, we can easily see that the following
corollary holds.

\begin{corollary}%
	\label{cor:invalid}
	If the public key has no  corresponding secret key,
	the success probability of any prover
	to be accepted by the verifier in all $\ell$ rounds
	of a full identification at most $1/2^\ell$.
\end{corollary}

The security of our scheme against key-only impersonation attack,
i.e., soundness, is based on the hardness of decisional MinRank problem,
as stated by the following theorem.

\begin{theorem}\label{thm:soundness}
	Let $\mathcal{A}$ be an algorithm such that given the public key
	$pk$, it is accepted in all $\ell$ rounds of the full identification
	protocol with probability $\varepsilon_{\mathcal{A}}\geqq\frac{1}{2^{\ell}}$,
	where the probability is taken over the random coins of $\mathcal{A}$,
	the key generator, and the verifier. Then, we can construct an algorithm
	which $(t,\varepsilon)$-solves the decisional MinRank problem associated
	with the finite field $\mathbb{F}$ and $r,m,n\in\mathbb{N}$
	such that the following holds. \vspace{-5mm}

	\begin{align*}
		\varepsilon  =\varepsilon_{\mathcal{A}}-\frac{1}{2^{\ell}},
		\quad
		t            = t_{\mathcal{A}},
	\end{align*}
	where $t_{\mathcal{A}}$ is the maximum total time of $\mathcal{A}$
	interacting in one full identification protocol.

\end{theorem}

\begin{corollary}
	If the decisional MinRank problem is $(t,\varepsilon)$-hard, then
	the success probability of
	any adversary attempting to impersonate a prover
	without secret key within  $t$ time units
	is upper-bounded by $\varepsilon+1/2^\ell$.
\end{corollary}

\subsection{Zero-Knowledgeness}
The following theorem is to guarantee that no knowledge on the secret leaked by
communication with the prover.

\begin{theorem}[Zero-Knowledgeness]%
	\label{thm:zk}
	For any verifier $V$, there exists an algorithm $M$ which given input
	the public key $pk$, perfectly simulates the view of verifier with the
	same distribution as the view of $V$ engaging with the prover possessing
	$pk$ and the secret key $sk$.
\end{theorem}

\section{Proofs of Main Theorems }

\subsection{Proof of Theorem \ref{thm:completeness}\label{subsec:Proof-of-Completeness}}

It is sufficient to show that for each challenge $c\in\{0,1,2,3\}$,
$Z$ which is generated accordingly in the procedure of the prover
will satisfy all the corresponding checking equations on the verifier
side.

Let us check for each case of challenge.
\begin{description}
	\item[Case{ }$c=0$:]
		Since $Z_{0,0}=U_{0,0}$ and $Y_{0,0}=H(U_{0,0})$,
		it is obvious that $H(Z_{0,0})=H(U_{0,0})=Y_{0,0}$ holds. Similarly,
		since $Z_{0,1}=U_{0,1}$ and $Y_{0,1}=H(U_{0,1})$ holds, it is obvious
		that $H(Z_{0,1})=H(U_{0,1})=Y_{0,1}$. Since $Z_{0,0}=U_{0,0}$ and
		$Z_{0,1}=U_{0,1}$, the followings hold.

		\begin{align*}
			\mathrm{rank}\left(Z_{0,1}-Z_{0,0}\right)
			 & =\mathrm{rank}\left(U_{0,1}-U_{0,0}\right) \\
			 & =\mathrm{rank}\left(T_{0}MS_{0}\right)     \\
			 & \stackrel{(a)}{=}\mathrm{rank}(M) = r.
		\end{align*}
		Eq. $(a)$ holds since $T_{0}$ and $S_{0}$ are invertible matrices.
		Since $Z_{1,0}=R_1$ and $Y_{1,2}=H(R_1)$, it is obvious
		that $H(Z_{1,0})=H(R_1)=Y_{1,2}$ holds.
		Also, we can easily see that $(\widetilde{S},
			\widetilde{T},\widetilde{X})=(S_1,T_1,X_1)$.
		Since $S_1, T_1$ are invertible matrices,
		so are $\widetilde{S}, \widetilde{T}$.
		Since $Z_{1,1}=\beta_1$, it is obvious that
		$(
			\widetilde{\gamma}_1,\ldots,\widetilde{\gamma}_{m-1}
			)=
			(
			\beta_{1,1},\ldots,\beta_{1,m-1}.
			)
		$
		Thus, the followings hold.
		\begin{align*}
			\widetilde{U}
			 & =\widetilde{T}
			\left(
			\sum_{i=1}^{m-1}\widetilde{\gamma}_{i}M_{i}
			\right)
			\widetilde{S}+\widetilde{X}                                  \\
			 & = T_1\left(\sum_{i=1}^{m-1}\beta_{1,i}M_{i}\right)S_1+X_1 \\
			 & = T_1 N_1 S_1+X_1
			= U_{1,0}.
		\end{align*}%
		Hence, since $Y_{1,0}=H(U_{1,0})$,
		$H(\widetilde{U})=H(U_{1,0})=Y_{1,0}$ holds.
	\item [{Case{ }$c=1$:}]%
	      Since $Z_{0,0}=R_0$ and $Y_{0,2}=H(R_0)$, it is obvious that
	      $H(Z_{0,0})=Y_{0,2}$ holds.
	      Hence, one can see that
	      $(\widehat{S},\widehat{T},\widehat{X})=
		      ({S}_0,{T}_0,{X}_0)$
	      holds.
	      Since $Z_{0,1}=\beta_0$, it is obvious that
	      $(\widehat{\gamma}_{1},\ldots,\widehat{\gamma}_{m-1})
		      =(\beta_{0,1},\ldots,\beta_{0,m-1})$ holds.
	      Thus, the following holds.
	      \begin{align*}
		      \widehat{U}
		       & =\widehat{T}
		      \left(
		      \sum_{i=1}^{m-1}\widehat{\gamma}_{i}M_{i}
		      \right)
		      \widehat{S}+\widehat{X}                                      \\
		       & = T_0\left(\sum_{i=1}^{m-1}\beta_{0,i}M_{i}\right)S_0+X_0 \\
		       & = T_0 N_0 S_0+X_0
		      = U_{0,0}.
	      \end{align*}%
	      Hence, since $Y_{0,0}=H(U_{0,0})$, automatically
	      $H(\widehat{U})=H(U_{0,0})=Y_{0,0}$ holds.
	      Next, since $Z_{1,0}=R_1$ and $Y_{1,2}=H(R_1)$, it is obvious that
	      $H(Z_{1,0})=Y_{1,2}$ holds.
	      Hence, one can see that
	      $(\widetilde{S},\widetilde{T},\widetilde{X})=
		      ({S}_1,{T}_1,{X}_1)$
	      holds.
	      Since $Z_{1,1}=\beta_1+\alpha$, it is obvious that
	      $(\widetilde{\mu}_{1},\ldots,\widetilde{\mu}_{m-1})
		      =(\beta_{1,1}+\alpha_1,\ldots,\beta_{1,m-1}+\alpha_{m-1})$ holds.
	      Thus, the following holds.
	      \begin{align*}
		      \widetilde{W}-\widetilde{T}M_0\widetilde{S}
		       & =\widetilde{T}
		      \left(
		      \sum_{i=1}^{m-1}\widetilde{\mu}_{i}M_{i}
		      \right)
		      \widetilde{S}+\widetilde{X}
		      -\widetilde{T}M_0\widetilde{S}
		      \\
		       & = T_1 \left(
		      \sum_{i=1}^{m-1}(\beta_{1,i}+\alpha_i)M_{i}
		      \right)S_1+X_1
		      \\
		       & \quad
		      - T_1 M_0 S_1                                   \\
		       & = T_1 N_1 S_1 + T_1 M S_1 + T_1 M_0 S_1 +X_1 \\
		       & \quad - T_1 M_0 S_1                          \\
		       & = T_1 N_1 S_1 + X_1
		      = U_{1,0}
	      \end{align*}%
	      Hence, since $Y_{1,0}=H(U_{1,0})$, automatically
	      $H(\widetilde{W}-\widetilde{T}M_0\widetilde{S})=H(U_{1,0})=Y_{1,0}$
	      holds.

	\item [{Case{ }$c=2$:}]%
	      This case is similar to the case $c=1$ with additional notes as follows:
	      \begin{itemize}
		      \item any variable in the form of $\widehat{*}$ notation
		            switches  with the resembling variable
		            in the form of $\widetilde{*}$ notation,
		      \item any variable in the form of  ${*}_0$ notation
		            switches  with the resembling variable
		            in the form of  ${*}_1$ notation,
		      \item for any numeric $j$, any variable in the form of  $*_{0,j}$ notation
		            switches  with the resembling variable in the form of
		            $*_{1,j}$ notation.
	      \end{itemize}

	\item [{Case{ }$c=3$:}]%
	      This case is similar to the case $c=0$ with the same additional
	      notes as in the case $c=2$.
\end{description}

\subsection{Proof of Proposition \ref{prop:three-combinations}\label{subsec:Proof-of-Three-Combinations}}

It is sufficient to show that from $Y$ and any combination of three
elements from the set of the valid responses $\{Z^{(0)},Z^{(1)},Z^{(2)},Z^{(3)}\}$,
we can compute $\alpha=(\alpha_{1},\ldots,\alpha_{m-1})\in\mathbb{F}^{m-1}$
such that $\mathrm{rank}\left(\sum_{i=1}^{m-1}\alpha_{i}M_{i}-M_{0}\right)=r$
holds, where $r$ and $M_{0},\ldots,M_{m-1}$ are generated by the
key generation algorithm as elements of the public key.

\begin{remark}
	Note that
	in our proposed scheme, we assume that $H$ has  computational
	binding property.
	Hence, we can assume that for any polynomial time algorithm,
	if $H(a)=H(b)$, then $a=b$ must hold
	except with negligible probability.
\end{remark}

\paragraph*{Case 1: $Y$ and $(Z^{(0)},Z^{(1)},Z^{(2)})$. }

Let $Z_{0,0}^{(1)}$ be parsed as $Z_{0,0}^{(1)}=(\widehat{S}^{(1)},\widehat{T}^{(1)},\widehat{X}^{(1)})$
and $Z_{0,1}^{(1)}$ be parsed as $Z_{0,1}^{(1)}=(\widehat{\gamma}_{1},\ldots\widehat{\gamma}_{m-1})$.
Also let $Z_{0,0}^{(2)}$ be parsed as $Z_{0,0}^{(2)}=(\widehat{S}^{(2)},\widehat{T}^{(2)},\widehat{X}^{(2)})$
and $Z_{0,1}^{(2)}$ be parsed as $Z_{0,1}^{(2)}=(\widehat{\mu}_{1},\ldots\widehat{\mu}_{m-1})$.
Since the following holds:

\begin{align*}
	H\left(\widehat{S}^{(1)},\widehat{T}^{(1)},\widehat{X}^{(1)}\right) & =H\left(\widehat{S}^{(2)},\widehat{T}^{(2)},\widehat{X}^{(2)}\right)=Y_{0,2},
\end{align*}
we can define as follows: $(\widehat{S},\widehat{T},\widehat{X}):=(\widehat{S}^{(1)},\widehat{T}^{(1)},\widehat{X}^{(1)})=(\widehat{S}^{(2)},\widehat{T}^{(2)},\widehat{X}^{(2)})$.
From $H(Z_{0,0}^{(0)})=Y_{0,0}$ and Eq. \eqref{eq:widehat-U}, we obtain as follows.
\begin{align}
	Y_{0,0} & =H(Z_{0,0}^{(0)})
	= H\left(\widehat{T}\left(\sum_{i=1}^{m-1}\widehat{\gamma}_{i}M_{i}							\right)
	\widehat{S}+\widehat{X}
	\right)
	\label{eq:y00-z00(0)-u(1)}                                                  \\
	        & \quad \Rightarrow Z_{0,0}^{(0)}=\widehat{T}\left(\sum_{i=1}^{m-1}
	\widehat{\gamma}_{i}M_{i}\right)\widehat{S}+\widehat{X},
	\nonumber
\end{align}
Similarly, from $H(Z_{0,1}^{(0)})=Y_{0,1}$ and Eq. \eqref{eq:widehat-W},
we also have the followings hold.
\begin{align}
	Y_{0,1}
	 & =H(Z_{0,1}^{(0)})
	\nonumber                                                                                                                                                \\
	 &
	=
	H\left(\widehat{T}\left(\sum_{i=1}^{m-1}\widehat{\mu}_{i}M_{i}
	\right)\widehat{S}+\widehat{X}-\widehat{T}M_{0}\widehat{S}\right)                                                                                        \\
	 & \Rightarrow Z_{0,1}^{(0)}=\widehat{T}\left(\sum_{i=1}^{m-1}\widehat{\mu}_{i}M_{i}\right)\widehat{S}+\widehat{X}-\widehat{T}M_{0}\widehat{S} \nonumber
\end{align}
Finally, we have the followings hold.
\begin{align*}
	\mathrm{rank}(Z_{0,1}^{(0)}-Z_{0,0}^{(0)}) & =\mathrm{rank}\left(\widehat{T}\left(\sum_{i=1}^{m-1}(\widehat{\mu}_{i}-\widehat{\gamma}_{i})M_{i}-M_{0}\right)\widehat{S}\right) \\
	                                           & \stackrel{(a)}{=}\mathrm{rank}\left(\sum_{i=1}^{m-1}(\widehat{\mu}_{i}-\widehat{\gamma}_{i})M_{i}-M_{0}\right),
\end{align*}
where
Eq. $(a)$ holds since $\widehat{S},\widehat{T}$ are non-singular.
Therefore, we can set $\alpha_{i}=\widehat{\mu}_{i}-\widehat{\gamma}_{i}$
for $i\in[1,m-1]$, since
$\mathrm{rank}(Z_{0,1}^{(0)}-Z_{0,0}^{(0)})=r$ holds.

\paragraph*{Case 2: $Y$ and $(Z^{(0)},Z^{(2)},Z^{(3)})$. }
Similar to Case 1. The only diffference is that all relations
and components of $Z^{(1)}$ in Case
1 are substituted by those of $Z^{(3)}$.

\paragraph*{Case 3: $Y$ and $(Z^{(1)},Z^{(2)},Z^{(3)})$. }
Let $Z_{1,0}^{(2)}$ be parsed as $Z_{1,0}^{(2)}=(\widetilde{S}^{(2)},\widetilde{T}^{(2)},\widetilde{X}^{(2)})$
and $Z_{1,1}^{(2)}$ be parsed as $Z_{1,1}^{(2)}=(\widetilde{\gamma}_{1},\ldots\widetilde{\gamma}_{m-1})$.
Also let $Z_{1,0}^{(1)}$ be parsed as $Z_{1,0}^{(1)}=(\widetilde{S}^{(1)},\widetilde{T}^{(1)},\widetilde{X}^{(1)})$
and $Z_{1,1}^{(1)}$ be parsed as $Z_{1,1}^{(1)}=(\widetilde{\mu}_{1},\ldots\widetilde{\mu}_{m-1})$.
Since the following holds:

\begin{align*}
	H\left(\widetilde{S}^{(1)},\widetilde{T}^{(1)},\widetilde{X}^{(1)}\right) & =H\left(\widetilde{S}^{(2)},\widetilde{T}^{(2)},\widetilde{X}^{(2)}\right)=Y_{1,2},
\end{align*}
we can define as follows: $(\widetilde{S},\widetilde{T},\widetilde{X}):=(\widetilde{S}^{(1)},\widetilde{T}^{(1)},\widetilde{X}^{(1)})=(\widetilde{S}^{(2)},\widetilde{T}^{(2)},\widetilde{X}^{(2)})$.
From $H(Z_{1,0}^{(3)})=Y_{1,0}$ and Eq. \eqref{eq:widetilde-U}, we obtain as follows.
\begin{align}
	Y_{1,0} & =H(Z_{1,0}^{(3)})
	= H\left(\widetilde{T}\left(\sum_{i=1}^{m-1}\widetilde{\gamma}_{i}M_{i}							\right)
	\widetilde{S}+\widetilde{X}
	\right)
	\\
	        & \quad \Rightarrow Z_{1,0}^{(3)}=\widetilde{T}\left(\sum_{i=1}^{m-1}
	\widetilde{\gamma}_{i}M_{i}\right)\widetilde{S}+\widetilde{X},
	\nonumber
\end{align}
Similarly, from $H(Z_{1,1}^{(3)})=Y_{1,1}$ and Eq. \eqref{eq:widetilde-W},
we also have the followings hold.
\begin{align}
	Y_{1,1}
	 & =H(Z_{1,1}^{(3)}) \nonumber                                                                                                                                       \\
	 &
	=
	H\left(\widetilde{T}\left(\sum_{i=1}^{m-1}\widetilde{\mu}_{i}M_{i}
	\right)\widetilde{S}+\widetilde{X}-\widetilde{T}M_{0}\widetilde{S}\right)                                                                                            \\
	 & \Rightarrow Z_{1,1}^{(3)}=\widetilde{T}\left(\sum_{i=1}^{m-1}\widetilde{\mu}_{i}M_{i}\right)\widetilde{S}+\widetilde{X}-\widetilde{T}M_{0}\widetilde{S} \nonumber
\end{align}
Finally, we have the followings hold.
\begin{align*}
	\mathrm{rank}(Z_{1,1}^{(3)}-Z_{1,0}^{(3)}) & =\mathrm{rank}\left(\widetilde{T}\left(\sum_{i=1}^{m-1}(\widetilde{\mu}_{i}-\widetilde{\gamma}_{i})M_{i}-M_{0}\right)\widetilde{S}\right) \\
	                                           & \stackrel{(a)}{=}\mathrm{rank}\left(\sum_{i=1}^{m-1}(\widetilde{\mu}_{i}-\widetilde{\gamma}_{i})M_{i}-M_{0}\right),
\end{align*}
where
Eq. $(a)$ holds since $\widetilde{S},\widetilde{T}$ are non-singular.
Therefore, we can set $\alpha_{i}=\widetilde{\mu}_{i}-\widetilde{\gamma}_{i}$
for $i\in[1,m-1]$, since
$\mathrm{rank}(Z_{1,1}^{(3)}-Z_{1,0}^{(3)})=r$ holds.

\paragraph*{Case 4: $Y$ and $(Z^{(0)},Z^{(1)},Z^{(3)})$. }
Similar to Case 3. The only diffference is that all relations
and components of $Z^{(2)}$ in Case
1 are substituted by those of $Z^{(0)}$.

\subsection{Proof Sketch of Corollary \ref{cor:invalid}}
%
Recall that based on Proposition \ref{prop:three-combinations},
we know that in any single round, if the prover can answer correctly
three out of four possible challenges from the verifier,
it means that the prover knows the secret key corresponding
public key. Thus, in the case that the public key has no
corresponding valid secret key, even a prover with unbounded
resources must not be able to answer correctly more than
two out of four possible challenges in any single round.
Otherwise, it will contradict with the assumption that
the public key
that the public key has no corresponding secret key.

\subsection{Proof Sketch of Theorem \ref{thm:soundness}}
Let define algorithm $\mathcal{D^{\mathsf{InputGen}}}(\mathbb{F},r,m,n)$
as follows. First, $\mathcal{D}$ retrieves inputs from the oracle
$\mathsf{InputGen}$ in the form of $m$ $n$-square matrices over
the finite field $\mathbb{F}$: $M_{0},\ldots,M_{m-1}$. Then, $\mathcal{D}$
simulates the key generation algorithm of the identification scheme
by setting the public key $pk$ as $r$ and $M_{0},\ldots,M_{m-1}$.
Next, $\mathcal{D}$ inputs $pk$ to $\mathcal{A}$ and runs $\mathcal{A}$
as the prover and $\mathcal{D}$ acts as the honest verifier. If $\mathcal{A}$
successfully gives valid responses in all $\ell$ rounds of the full
identification protocol, $\mathcal{D}$ outputs $1$, otherwise, $\mathcal{D}$
outputs $0$.
Note that if $\mathsf{InputGen}$ is $\mathsf{IGen}$, the probability of $\mathcal{D}$
outputs $1$ is exactly $\varepsilon_\mathcal{A}$.
Meanwhile, when $\mathsf{InputGen}$ is $\mathsf{LossyGen}$,
based on Corollary \ref{cor:invalid}, the probability of
$\mathsf{D}$ outputs $1$ is at most $1/2^\ell$.
Thus, denoting the system parameters $(\mathbb{F},r,m,n)$ as $\mathrm{par}$,
we obtain as follows.
\begin{align*}
	\left|\Pr[\mathcal{D}^{\mathsf{IGen}}(\mathrm{par})=1]
	-\Pr[\mathcal{D}^{\mathsf{LossyGen}}(\mathrm{par})=1] \right |
	 & \geqq \varepsilon_\mathcal{A}- \frac{1}{2^\ell}.
\end{align*}
This proves  Theorem \ref{thm:soundness}.

\subsection{Proof Idea of Theorem \ref{thm:zk}}
It is sufficient to prove that
given any $c\in\{0,1,2,3\}$, we can
create valid response  $Z_{0,0},Z_{0,1},Z_{1,0},Z_{1,1}$
and the commitment $Y_0,Y_1$ without using secret key
such that their distribution is the same as
the distribution of the response and commitment
generated by a honest prover who possesses
valid secret key. Note that  we can put the
responses and commitment
into two independent groups:
$(Y_0,Z_{0,0},Z_{0,1})$ and
$(Y_1,Z_{1,0},Z_{1,1})$,
such that
each group is corresponding to the
set of responses and commitment
in Courtois'  ID scheme~\cite{DBLP:conf/asiacrypt/Courtois01}.
Hence, it is easy to see that
we can apply the proof of zero-knowledge
for Courtois' ID scheme into our proposed
scheme.

\section{Parameter Selections}
\subsection{Complexity of MinRank Problem}
In this subsection, we review known attacks against MinRank Problem to select some practical parameters.

There are two types of attack.
First one is to mainly use linear algebra and second one is to reduce the MinRank problem into an MP problem.
Set $\mathbb{F}=\mathbb{F}_q$.
\\

\paragraph*{Linear algebra type}
There exist 4 attacks in this type.
Our review for this type mainly follows the Subsection 4.2 in \cite{DBLP:conf/asiacrypt/Courtois01}.
\\

\noindent
(i) {\bf Exhaustive search attack}:
This attack is to find $M:=\sum_{i=1}^{m-1}\alpha_iM_i-M_0$ or a matrix with rank $\leq r$ from the linear combinations of $M_0,\dots,M_{m-1}$.
The complexity to find $M$ from $M_0,\dots,M_{m-1}$ is given by
\[
q^{m-1}(r+1)^{\omega},
\]
where $2<\omega \leq 3$ is a linear algebra constant. 

Next, consider the complexity to find a matrix with rank $\leq r$.
The probability that a square matrix with size $n$ is of rank $\ell$ is given by
\[
P(n,\ell):=\frac{(q^n-1)^2(q^{n}-q)^2\cdots(q^{n}-q^{\ell-1})^2}{(q^{\ell}-1)\cdots (q^{\ell}-q^{\ell-1})\cdot q^{n^2}}.
\]
We assume that the probability that a linear combination of $M_0,\cdots,M_{m-1}$ is of rank $\ell$ 
is $P(n,\ell)$. 
Then the complexity to find  a matrix with rank $\leq r$ from the linear combinations of $M_0,\dots,M_{m-1}$ 
is given by
\[
\left(\sum_{\ell=1}^{r}P(n,\ell)\right)^{-1}(r+1)^{\omega}.
\]

\noindent
(ii) {\bf Kernel attack}:
This attack is to find an element of the kernel of $M$.
The complexity is given by
\[
\text{Min}\left(q^{\lceil\frac{m}{n}\rceil r},q^{\lfloor\frac{m}{n}\rfloor r+(m \mod n))}\right)m^{\omega}.
\]

\noindent
(iii) {\bf``Big m" attack}: 
This attack is valid for $m\gg n$.
The complexity is given by
\[
q^{\text{Max}(0,n(n-r)-m)}\cdot(n(n-r))^{\omega}.
\]

\noindent
(iv) {\bf Syndrome attack}:
This attack is valid for $m\gg n$.
The complexity is given by
\[
q^{\text{Max}(\frac{n^2-m-1}{2},nr-m-\frac{r^2}{4})}\cdot\mathcal{O}(n^2r).
\]

There is another attack using submatrices that works under $r\ll n$ (see also \cite{DBLP:conf/asiacrypt/Courtois01}).
However, in our setting, we will choose the rank $r$ to be about $n/2$.
Therefore, we skip such an attack.
\\

\paragraph*{MP type}
The MinRank problem can be reduced to the problem that solves a system of polynomial equations (namely, MP problem).
There exist three attacks in this type: (v) Kipnis-Shamir attack, (vi) Minors modeling attack, and (vii) Support minors modeling attack.
\\

\noindent (v) {\bf Kipnis-Shamir attack} \cite{DBLP:conf/crypto/KipnisS99}:
Let $c$ be an integer such that $\lceil m/(n-r)\rceil \leq c\leq n-r$.
By considering $\alpha_1,\dots ,\alpha_m$ and kernel basis vectors $\{\bold{y}_1,\dots,\bold{y}_c\}$ of $\sum_{i=1}^m\alpha_iM_i-M_0$ as variables, Kipnis-Shamir attack solves the quadratic system consisting of
$
\bold{y}_i\cdot \left(\sum_{i=1}^m\alpha_iM_i-M_0\right)=0.
$
The complexity estimations of this attack are given as Table \ref{table:CmplKS}.
Here, for each estimation, we take $c$ giving the minimum value in Table \ref{table:CmplKS}. 

\begin{table}[h]
\caption{Complexity estimations for the Kipnis-Shamir attack}\label{table:CmplKS}
\centering
\begin{tabular}{|c|c|c|} \hline
Faugere et al. \cite{DBLP:conf/crypto/FaugereLP08}&Verbel et al. \cite{DBLP:conf/pqcrypto/VerbelBCPS19}&Nakamura et al. \cite{DBLP:journals/iacr/NakamuraWI20}\\ \hline \hline
$\log_2(q){n\choose r}^{\omega (n-r)}$&$\left (m{cr+D_{KS}-1\choose D_{KS}}\right )^\omega$&$ {m+cr+D_{mgd}\choose D_{mgd}}^\omega $\\ \hline
\end{tabular}
\end{table}

\noindent
Here, $D_{KS}$ is defined as follows.
Let $d_{KS}=\min _{1\leq d\leq r}\{d:{r\choose d}n>{r\choose d+1}m \}$. 
Then $D_{KS}=d_{KS}+2$.
Moreover, $D_{mgd}$ is defined as follows.
Set 
\begin{align*}
&\sum _{(e_0,e_1,\dots, e_c)\in \mathbb{Z}^{c+1}}a_{(e_0,e_1,\dots, e_c)}t_0^{e_0}t_1^{e_1}\cdots t_c^{e_c}\\
:=&\cfrac{\prod _{i=1}^c(1-t_0t_i)^n}{(1-t_0)^m(1-t_1)^r\cdots (1-t_c)^r}.
\end{align*}
Then define $\displaystyle D_{mgd}=\min \left \{\sum _{i=1}^ce_i: a_{(e_0,e_1,\dots ,e_c)}<0\right \}$.
\\

\noindent (vi) {\bf Minors modeling attack} \cite{DBLP:conf/issac/FaugereDS10}:
This attack solves the system consisting of the $(r+1)$-minors of $\sum_{i=1}^m\alpha_iM_i-M_0$, whose variables are $\alpha_1,\dots ,\alpha_m$.
The complexity is estimated by ${m+r\choose r}^\omega $.\\

\noindent (vii) {\bf Support Minors modeling attack} \cite{DBLP:conf/asiacrypt/BardetBCGPSTV20}:
This attack solves a quadratic system whose variables are $\alpha_1,\dots ,\alpha_m$ and $r$-minors, and its complexity is estimated by 
\[
3{m+D_{Spp}\choose D_{Spp}}^2{n\choose r}^2(r+1)m.
\]
Here, $D_{Spp}$ is defined as follows.
For $b\ge1$, set $R_{m,n,r}(b)=\sum _{i=1}^b(-1)^{i+1}{n\choose r+i}{n+i-1\choose i}{m+b-i-1\choose b-i}$ and $\mathcal{M}(b,1)={m+b\choose b}{n\choose r}$.
Then define $D_{Spp}=\min \{b\mid R_{m,n,r}(b)>\mathcal{M}(b,1)-1\}$.

In this subsection, we review known attacks against MinRank Problem to select some practical parameters.

There are two types of attack.
First one is to mainly use linear algebra and second one is to reduce the MinRank problem into the problem that solves a system of polynomial equations (namely, MP problem).
Set $\mathbb{F}=\mathbb{F}_q$, and let $\omega $ be a linear algebra constant.

\paragraph*{Linear algebra type}
There exist 4 attacks in this type, and Table \ref{table:attackLA} lists these complexity estimations by according to the Subsection 4.2 in \cite{DBLP:conf/asiacrypt/Courtois01}.

\begin{table}[h]
\caption{Complexity estimations for attacks of Linear algebra type}\label{table:attackLA}
\centering
\begin{tabular}{|r|c|} \hline
Attack & Complexity Estimation\\ \hline \hline 
Exhaustive &  $\left(\sum_{\ell=1}^{r}P(n,\ell)\right)^{-1}(r+1)^{\omega}$\\ \hline 
Kernel&$\text{Min}\left(q^{\lceil\frac{m}{n}\rceil r},q^{\lfloor\frac{m}{n}\rfloor r+(m \mod n))}\right)m^{\omega}$\\ \hline 
Big-m&$q^{\text{Max}(0,n(n-r)-m)}\cdot(n(n-r))^{\omega}$\\ \hline  
Syndrome &$q^{\text{Max}(\frac{n^2-m-1}{2},nr-m-\frac{r^2}{4})}\cdot\mathcal{O}(n^2r)$\\ \hline 
\end{tabular}
\end{table}

\noindent Here, in Table \ref{table:attackLA}, we used as 
\[
P(n,\ell):=\frac{(q^n-1)^2(q^{n}-q)^2\cdots(q^{n}-q^{\ell-1})^2}{(q^{\ell}-1)\cdots (q^{\ell}-q^{\ell-1})\cdot q^{n^2}}.
\]
There is another attack using submatrices that works under $r\ll n$ (see also \cite{DBLP:conf/asiacrypt/Courtois01}).
However, in our setting, we will choose the rank $r$ to be about $n/2$.
Therefore, we skip such an attack.

\paragraph*{MP type}
There exist three attacks in this type, and these complexity estimations are given as in Table \ref{table:attackMP}:
\begin{table}[h]
\caption{Complexity estimations for attacks of MP type}\label{table:attackMP}
\centering
\begin{tabular}{|r|c|} \hline
Attack & Complexity Estimation\\ \hline \hline 
Kipnis-Shamir & $\log_2(q){n\choose r}^{\omega (n-r)}$\cite{DBLP:conf/crypto/FaugereLP08},\\ 
&$\left (m{cr+D_{KS}-1\choose D_{KS}}\right )^\omega$\cite{DBLP:conf/pqcrypto/VerbelBCPS19}, or \\ 
& $ {m+cr+D_{mgd}\choose D_{mgd}}^\omega $\cite{DBLP:journals/iacr/NakamuraWI20}\\ \hline
Minors modeling & ${m+r\choose r}^\omega $\cite{DBLP:conf/issac/FaugereDS10}\\ \hline 
Support minors modeling & $3{m+D_{Spp}\choose D_{Spp}}^2{n\choose r}^2(r+1)m$\cite{DBLP:conf/asiacrypt/BardetBCGPSTV20}\\ \hline 
\end{tabular}
\end{table}

\noindent Here we took $c$ giving the minimum value in Table \ref{table:attackMP} for $\lceil m/(n-r)\rceil \leq c\leq n-r$, and used the following notations:
\begin{itemize}
\item $D_{KS}:=\min _{1\leq d\leq r}\{d:{r\choose d}n>{r\choose d+1}m \}+2$
\item $D_{mgd}:=\min \left \{\sum _{i=1}^ce_i: a_{(e_0,e_1,\dots ,e_c)}<0\right \} $ where 
\begin{align*}
    &\sum _{{\bf e}=(e_0,\dots, e_c)\in \mathbb{Z}^{c+1}}\hspace{-20pt}a_{{\bf e}}\,t_0^{e_0}\cdots t_c^{e_c}\\
    &\quad \quad\quad\quad :=\cfrac{\prod _{i=1}^c(1-t_0t_i)^n}{(1-t_0)^m(1-t_1)^r\cdots (1-t_c)^r}.
\end{align*}
\item  $D_{Spp}:=\min \{b\mid R_{m,n,r}(b)>\mathcal{M}(b,1)-1\}$ where $R_{m,n,r}(b)=\sum _{i=1}^b(-1)^{i+1}{n\choose r+i}{n+i-1\choose i}{m+b-i-1\choose b-i}$ and $\mathcal{M}(b,1)={m+b\choose b}{n\choose r}$ for $b\geq 1$.
\end{itemize}


\subsection{Communication Costs}

We will estimate the communication costs based on the assumption that
we use random seed and pseudorandom generator to generates $S_{0},S_{1},T_{0},T_{1},X_{0},X_{1},\beta_{0},\beta_{1}$.

Let $Z^{(c)}$ denote the valid response of the prover with respect to challenge
$c$ for any $c\in\{0,1,2,3\}$. For simplicity, here we assume that
all matrices are $n$-square matrices and $\mathbb{F=\mathbb{F}}_{q}$,
where $q$ is a power of some prime. Thus, we have as follows.

\begin{align*}
	|Z^{(0)}|=|Z^{(3)}| & \approx2n^{2}\log_{2}q+|\mathrm{seed}_{\overline{STX}}|+|\mathrm{seed}_{\beta}|, \\
	|Z^{(1)}|=|Z^{(2)}| & \approx2|\mathrm{seed}_{\overline{STX}}|+|\mathrm{seed}_{\beta}|+(m-1)\log_{2}q,
\end{align*}
where $\mathrm{seed}_{\overline{STX}}$ is the seed for generating
$(S_{0},T_{0},X_{0})$ or $(S_{1},T_{1},X_{1})$ and $\mathrm{seed}_{\beta}$
is the seed for generating $\beta_{0}$ or $\beta_{1}$.

Let $\#R_{1/2}$ and $\#R_{2/3}$ denote
the necessary number of rounds to achieve $\ell$-bit
security for our scheme and Curtois' respectively.  
It is easy to see that
$\#R_{1/2}=\ell$ and $\#R_{2/3}=2/3\lceil\ell/(\log_2{3}-1)\rceil$.
Assuming that the seeds for $\ell$-bit security are $\ell$ bits,
we have the following general equations for estimating the average 
total
communication costs for $\ell$-bit security.

\begin{align*}
	\#Z_{1/2} & \approx
	\ell 
	\left(
	\left(
	n^2 + (m-1)
	\right)\log_2 q
	+ \frac{5\ell}{2}
	\right),            \\
	\#Z_{2/3} & \approx
	\frac{2}{3}
	\left\lceil
	\frac{\ell}{\log_2 3-1}
	\right\rceil
	\left(
	\left(
	n^2 + (m-1)
	\right)\log_2 q
	+ \frac{3\ell}{2}
	\right),
\end{align*}
where $\#Z_{1/2}$ and  $\#Z_{2/3}$
denote the average total communication costs of our proposed scheme
and that of Curtois' respectively.

\subsection{Security Parameters}

Based on various attacks on the MinRank
problem which we review above, 
we recommend parameters for $128$, $192$ and $256$-bit security with $q=2$
as follows. Here we denote bytes as B.

\begin{table}[htbp]
\begin{tabular}{ccccrr}
\hline
bit security & $(n,m,r)$ &  $\#R_{1/2}$ & $\#R_{2/3}$ & $\#Z_{1/2}$ 
             & $\#Z_{2/3}$ \\
\hline
$128$ & $(26,209,13)$ & $128$ & $146$ & $19264$ B & $19637$ B \\
$192$ & $(33,331,17)$ & $192$ & $220$ & $45576$ B & $46800$ B \\
$256$ & $(39,469,20)$ & $256$ & $292$ & $84128$ B & $86614$ B \\
\hline 
\end{tabular}
\end{table}

\section{Conclusion}
In this paper, we have shown a construction
of a new three-pass ID scheme with half cheating probability.
In practice, compared to Curtois' ID scheme \cite{DBLP:conf/asiacrypt/Courtois01},
our scheme requires less number of repetitions 
to achieve the desired security level and  
has less average total communication cost.
As a future work, we aim to construct a digital signature 
based on our proposed ID scheme and prove its security 
against quantum adversaries.

\vspace{1mm}

\noindent{\bf{Acknowledgements}}
This work was supported by JST CREST Grant Number
JPMJCR2113, JSPS KAKENHI Grant Number JP19K20266, JP20K19802,
JP20K03741, JP18H01438, and JP18K11292.

\bibliographystyle{plain}
\bibliography{Minrank}

\end{document}